\documentclass[journal,twoside,web]{IEEEtran}
\usepackage{cite}
\usepackage{dutchcal}       
\usepackage{amsmath,amssymb,amsfonts}
\usepackage{threeparttable }
\usepackage{mathrsfs}
\usepackage{algorithmic}    
\usepackage{hyperref}       
\usepackage{xurl}           
\usepackage{booktabs}       
\usepackage{amsfonts}       
\usepackage{nicefrac} 
\usepackage{microtype}      
\usepackage{lipsum}
\usepackage{graphicx}
\usepackage{textcomp}
\usepackage{float}
\usepackage{makecell}
\usepackage{multirow}
\usepackage{color}
\usepackage{diagbox}
\usepackage{hyperref}[colorlinks,linkcolor=black]
\usepackage{color,soul}
\usepackage{gensymb}
\usepackage{tablefootnote}
\usepackage[linesnumbered,ruled]{algorithm2e}

\def\etal{\emph{et al.}}

\def\etc{\emph{etc}}

\begin{document}
\title{AdLER: Adversarial Training with Label Error Rectification for One-Shot Medical Image Segmentation}
\author{Xiangyu~Zhao, Sheng~Wang, Zhiyun~Song, Zhenrong~Shen, Linlin~Yao, Haolei~Yuan, Qian~Wang, and Lichi~Zhang
\thanks{Corresponding author: L. Zhang (e-mail: lichizhang@sjtu.edu.cn).}
\thanks{X. Zhao, S. Wang, Z. Song, Z. Shen, L. Yao, and L. Zhang are with School of Biomedical Engineering, Shanghai Jiao Tong University, Shanghai 200030, China (e-mails: \{xiangyu.zhao, wsheng\}@sjtu.edu.cn, zhiyunsung@gmail.com, \{zhenrongshen, yaolinlin23, lichizhang\}@sjtu.edu.cn).}
\thanks{H. Yuan is with GeneScience Pharmaceutical Co., Ltd, Shanghai 200030, China (e-mail: yuanhaolei@hotmail.com).}
\thanks{Q. Wang is with School of Biomedical Engineering, ShanghaiTech University, Shanghai 201210, China, 
and with Shanghai Clinical Research and Trial Center, Shanghai 201210, China (e-mail: qianwang@shanghaitech.edu.cn).}
}
\maketitle
\bibliographystyle{ieeetr}
\begin{abstract}
Accurate automatic segmentation of medical images typically requires large datasets with high-quality annotations, making it less applicable in clinical settings due to limited training data. One-shot segmentation based on learned transformations (OSSLT) has shown promise when labeled data is extremely limited, typically including unsupervised deformable registration, data augmentation with learned registration, and segmentation learned from augmented data. However, current one-shot segmentation methods are challenged by limited data diversity during augmentation, and potential label errors caused by imperfect registration.
To address these issues, we propose a novel one-shot medical image segmentation method with adversarial training and label error rectification (AdLER), with the aim of improving the diversity of generated data and correcting label errors to enhance segmentation performance. Specifically, we implement a novel dual consistency constraint to ensure anatomy-aligned registration that lessens registration errors. Furthermore, we develop an adversarial training strategy to augment the atlas image, which ensures both generation diversity and segmentation robustness. We also propose to rectify potential label errors in the augmented atlas images by estimating segmentation uncertainty, which can compensate for the imperfect nature of deformable registration and improve segmentation authenticity.
Experiments on the CANDI and ABIDE datasets demonstrate that the proposed AdLER outperforms previous state-of-the-art methods by 0.7\% (CANDI), 3.6\% (ABIDE "seen"), and 4.9\% (ABIDE "unseen") in segmentation based on Dice scores, respectively.
The source code will be available at \url{https://github.com/hsiangyuzhao/AdLER}.
\end{abstract}


\section{Introduction}
\label{sec:introduction}
Automatic medical image segmentation aims to efficiently delineate various anatomical structures and regions of interest from medical images, which plays a fundamental role in numerous clinical applications \cite{kaur2017survey}. The development of deep learning has led to extensive research on fully supervised medical image segmentation and has yielded promising results \cite{ronneberger2015u, isensee2021nnu, chen2021transunet, cao2021swin}. However, these methods are typically designed for specific tasks with corresponding large datasets with labeled images for modeling training, which poses a significant challenge in clinical practice due to the high cost of preparations for medical images with manual annotations, and with the issue of generalization to other tasks. 

Many attempts have been made to resolve the aforementioned issue, and few-shot learning \cite{ouyang2022self, tang2021recurrent} has gained much attention that can significantly alleviate the burden of abundant label annotations. However, they still require some labeled images which in some scenarios remains inconvenient or even impractical to fulfill. The recent introduction of the Segment Anything Model (SAM) \cite{kirillov2023segment} based on large-scale foundation model has presented its great capability as a universal and zero-shot segmentation tool to obtain accurate labeling without the pre-requisite of the training dataset for the target task. However, several studies \cite{zhang_towards_2023} reported its limited capacity in the medical image segmentation, which is more challenging than natural images due to high structural complexity, low contrast with weak boundaries, and demanding requirements for clinical domain knowledge.   

\begin{figure*}
    \centering
    \includegraphics[width=0.90\textwidth]{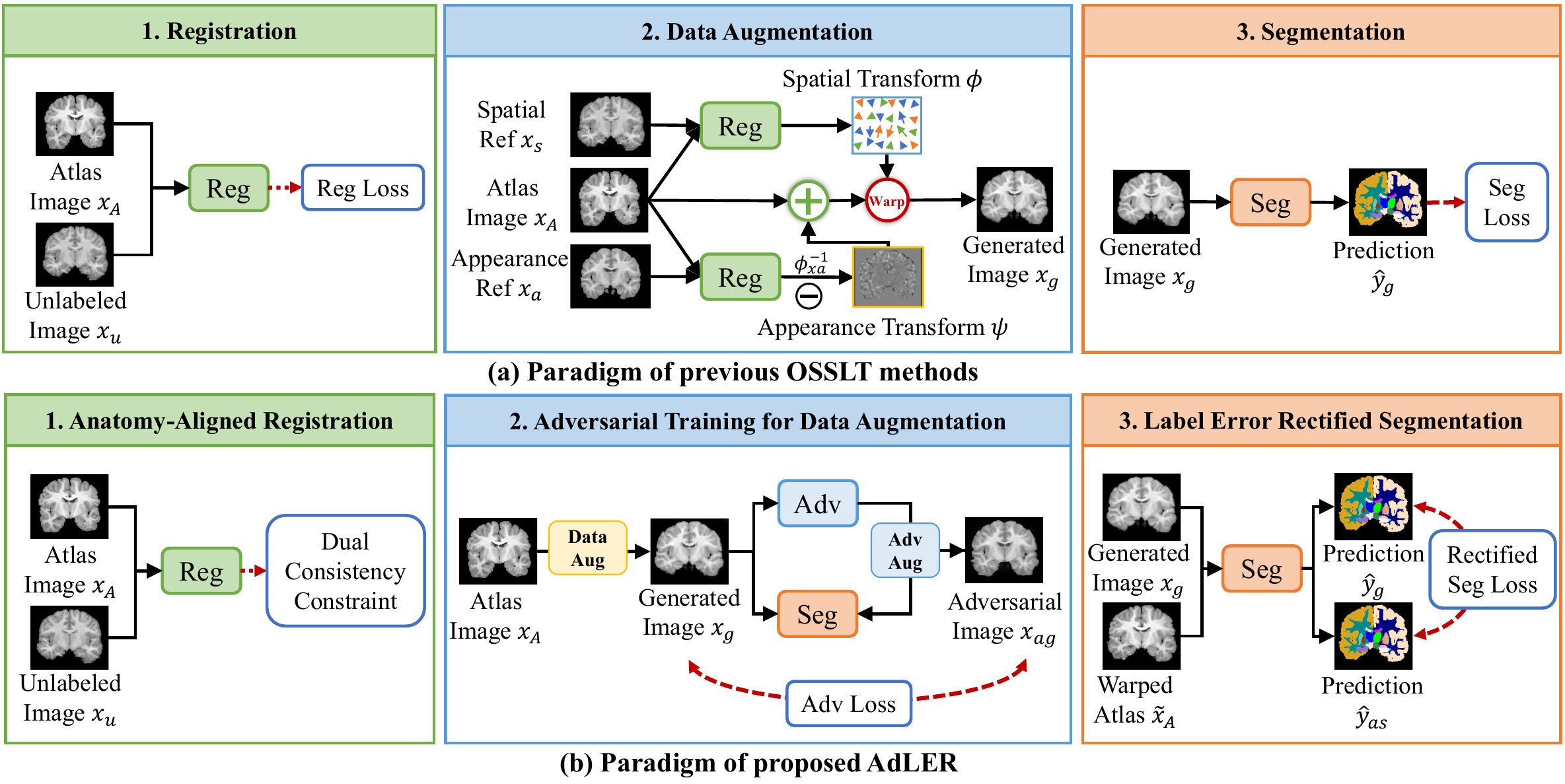}
    \caption{Illustration of one-shot segmentation based on learned transformations (OSSLT) and our proposed AdLER, which includes: (1) \emph{Anatomy-aligned registration} that achieves better anatomy alignment by proposed dual consistency constraint, which reduces the effects of label errors caused by imperfect registration; (2) \emph{Adversarial training} for data augmentation that improves both the diversity of augmentation and robustness of segmentation; (3) \emph{Label error rectified segmentation} that further rectifies potential label errors during segmentation to improve segmentation accuracy.}
    \label{fig:OSSLT}
\end{figure*}

One-shot segmentation based on learned transformations (OSSLT) has been developed to significantly reduce the dependence on labeled data in automatic segmentation \cite{zhao2019data, ding2021modeling, wang2020lt, he2022learning}. These methods typically utilize a single atlas image as the reference, which also exhibit similarities to the atlas-based segmentation techniques that were popular before the advent of deep learning \cite{collins1995automatic, lorenzo2002atlas, lotjonen2010fast, coupe2011patch}. The construction of OSSLT consists of three major steps: registration, data augmentation, and segmentation, as shown in Fig. \ref{fig:OSSLT} (a). The unsupervised deformable image registration is initially trained using unlabeled data to capture spatial and appearance diversity. After that, two reference images (spatial and appearance) are randomly selected from the unlabeled dataset to create a labeled image by applying the spatial and appearance transformations learned from registration to the atlas image and label, simulating the anatomy and appearance of the reference images. In this way, the data augmentation can produce synthesized labeled images with high diversity, which are utilized to train the supervised segmentation model.


However, current design of OSSLT methods faces two significant challenges that hinder their applications in clinical scenarios. The first challenge is the high dependency on the variety of unlabeled images available to generate augmented data with sufficient diversity, which is crucial to guarantee the quality of the training dataset for constructing the segmentation model. For example, Zhao \etal \cite{zhao2019data} proposed to augment the atlas image by sampling unlabeled images as spatial or appearance references, but it heavily needed a large number of unlabeled images to ensure diverse generated data. Another attempt proposed by Ding \etal \cite{ding2021modeling} involves constructing two variational autoencoders (VAEs) to generate spatial and appearance transformations respectively after the registration process. However, the performance of VAEs also depends on the implemented registration method. Similarly, He \etal \cite{he2022learning} directly sampled spatial and appearance transformations to introduce additional perturbations through data augmentation. Nevertheless, its sampling process relies on predefined rules, which limits the potential capacity of transformation sampling.

\begin{figure}
    \centering
    \includegraphics[width=0.45\textwidth]{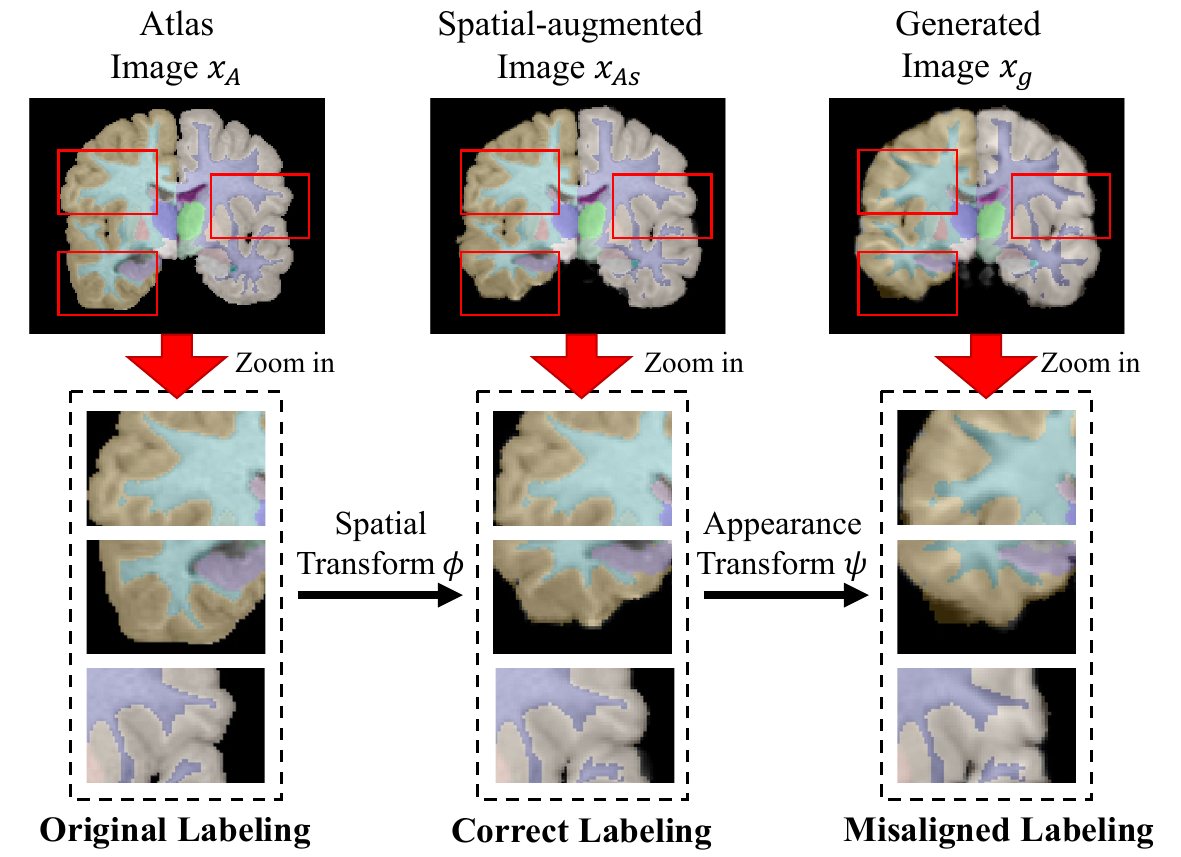}
    \caption{The illustration of potential label errors introduced by appearance transformation. Since the appearance transformation is only applied to the image, the label remains unchanged but the actual anatomical structure after applying the appearance transformation could alter, which leads to label errors in the augmented image.}
    \label{fig:error}
\end{figure}

Secondly, during the augmentation step, the augmented image and label may become misaligned due to the appearance transformation as illustrated in Fig. \ref{fig:error}. When spatial transformations are applied to the atlas image and label simultaneously, their correspondence should theoretically be maintained (as depicted in the middle column of Fig. \ref{fig:error}). However, when applying appearance transformation, the potential local changes in the tissue appearances can influence the anatomical structure of the augmented images, resulting in a mismatch between the augmented image and its label (as depicted in the right column of Fig. \ref{fig:error}). This is attributed to the appearance transformation in data augmentation that is highly dependent on the performance of deformable image registration. As deformable registration cannot achieve absolutely alignment of anatomical structures in two images, the resulting appearance transformation may therefore contain structural information that influences the actual semantic labels, in addition to intensity variations.  Furthermore, the presence of structural abnormalities in medical images such as lesions can also be included in the appearance transformations and affect the actual semantic labels.

To address the aforementioned issues, we propose a novel one-shot medical segmentation based on adversarial training and label error rectification (AdLER) as shown in Fig. \ref{fig:OSSLT} (b). (1) Since label errors arise from imperfect registration, we design a novel dual consistency constraint to ensure anatomy-aligned registration, which significantly improves topology reversibility \cite{holden2007review} essential to registration performance. (2) We introduce an adversarial training strategy to simultaneously improve the diversity of augmentation and the robustness of segmentation. Specifically, the image generated by data augmentation in common OSSLT methods is further augmented by an adversarial network to provide extra diversity and perturbations, while the segmentation is trained to minimize segmentation differences before and after adversarial augmentation. (3) We propose to rectify potential label errors by estimating segmentation uncertainty. Two augmented images, one augmented by spatial transformation only and the other by a combination of spatial and appearance transformation, are fed to the network. Since label errors are likely to be introduced by the appearance transformations, the segmentation difference between the two images indicates the potential label error regions, which are then rectified by uncertainty estimation. We validate our method on CANDI \cite{kennedy2012candishare} and ABIDE \cite{di2014autism} which are the two publicly available benchmark datasets. Our method achieves state-of-the-art performance in both datasets, demonstrating the superiority of the AdLER method.



\section{Related Works}
\subsection{Few-Shot Image Segmentation}
 Learning-based segmentation methods present a significant challenge in the presence of limited annotated data, and relevant attempts for few-shot segmentation in natural images are typically based on meta-learning techniques\cite{wang2020generalizing}. The objective of these methods \cite{dong2018few, wang2019panet, zhang2020sg, li2021adaptive, kwon2021dual, chen2021semantically, yang2021mining} is to expand segmentation from "seen" classes with substantial labeled data to "unseen" classes with limited labeled data. In inference, it can therefore be applied to segment novel class labels that do not exist during training. For example, Dong \etal \cite{dong2018few} proposed prototypical learning to address the few-shot segmentation task, which has established a baseline for the follow-up studies: Wang \etal \cite{wang2019panet} developed PANet to learn few-shot segmentation through prototype alignment, Kwon \etal \cite{kwon2021dual} combined contrastive learning with meta-learning to improve class separability in few-shot segmentation, and Chen \etal \cite{chen2021semantically} proposed to extract semantically meaningful classes to enhance the performance of few-shot segmentation. 

In addition to their applications in natural image segmentation, similar strategies have also been implemented in medical image segmentation. For instance, Ouyang \etal \cite{ouyang2022self} developed a self-supervision strategy for the few-shot medical image segmentation, and Tang \etal \cite{tang2021recurrent} introduced a recurrent mask refinement framework to improve the performance of medical image segmentation. However, the aforementioned approaches have several limitations when applied to medical images, as these methods require a considerable number of labeled images for training to ensure sufficient generalization ability to novel class labels, which is in contradiction with the medical image segmentation scenario where annotated data are usually limited. Furthermore, these methods are typically designed as 2D instead of 3D segmentation as they require a high-performing backbone for effective feature extraction\cite{chenrole}, while the availability of 3D pretrained backbones is quite limited. Consequently, the segmentation performance for 3D images such as CT and MRI is affected by the loss of inter-slice spatial information using these 2D segmentation methods. 

\subsection{One-Shot Segmentation Based on Learned Transformations}
The primary concept behind one-shot segmentation based on learned transformations (OSSLT) is to utilize unsupervised deformable registration to assist the supervised segmentation task. The intuition behind this idea originates from the inner connections between registration and segmentation. Many studies have investigated the combination of registration learning and segmentation learning. For instance, He \etal \cite{he2020deep} proposed a complementary joint model for registration and few-shot segmentation. Xu \etal \cite{xu2019deepatlas} developed DeepAtlas to implement registration and segmentation simultaneously. 
In contrast to few-shot segmentation approaches based on meta-learning, OSSLT is better suited for medical image segmentation, as they require very few labeled samples for model construction during both the training and inference stages, and most of these methods are designed in 3D segmentation. 

With the incorporation of deformable registration, OSSLT utilizes the atlas image to establish its transformation mappings to the target unlabeled images in an unsupervised manner. They can be subsequently transferred to the segmentation process through transformations in both spatial structure and image appearance.
For instance, Zhao \etal \cite{zhao2019data} proposed to implement separate spatial and appearance transformations through registration, and sample unlabeled images as reference images for data augmentation. Ding \etal \cite{ding2021modeling} explored differently by modeling the unlabeled data distribution with two separate VAEs to generate appearance and spatial transformations, which is followed by data augmentation and segmentation training. Wang \etal \cite{wang2020lt} proposed LT-Net which introduces forward-backward consistency to enhance registration performance, and generate improved spatial transformations for data augmentation. Finally, He \etal \cite{he2022learning} introduced BRBS with the aim of improving the authenticity, diversity, and robustness of the OSSLT method. They also make use of the spatial prior inherent in medical images. However, as discussed in Section \ref{sec:introduction}, ensuring the performance of OSSLT-based segmentation remains challenging due to the diversity demands of the generated data, and the potential label errors during atlas augmentation caused by the appearance transformation.

\begin{figure*}[t]
    \centering
    \includegraphics[width=0.90\textwidth]{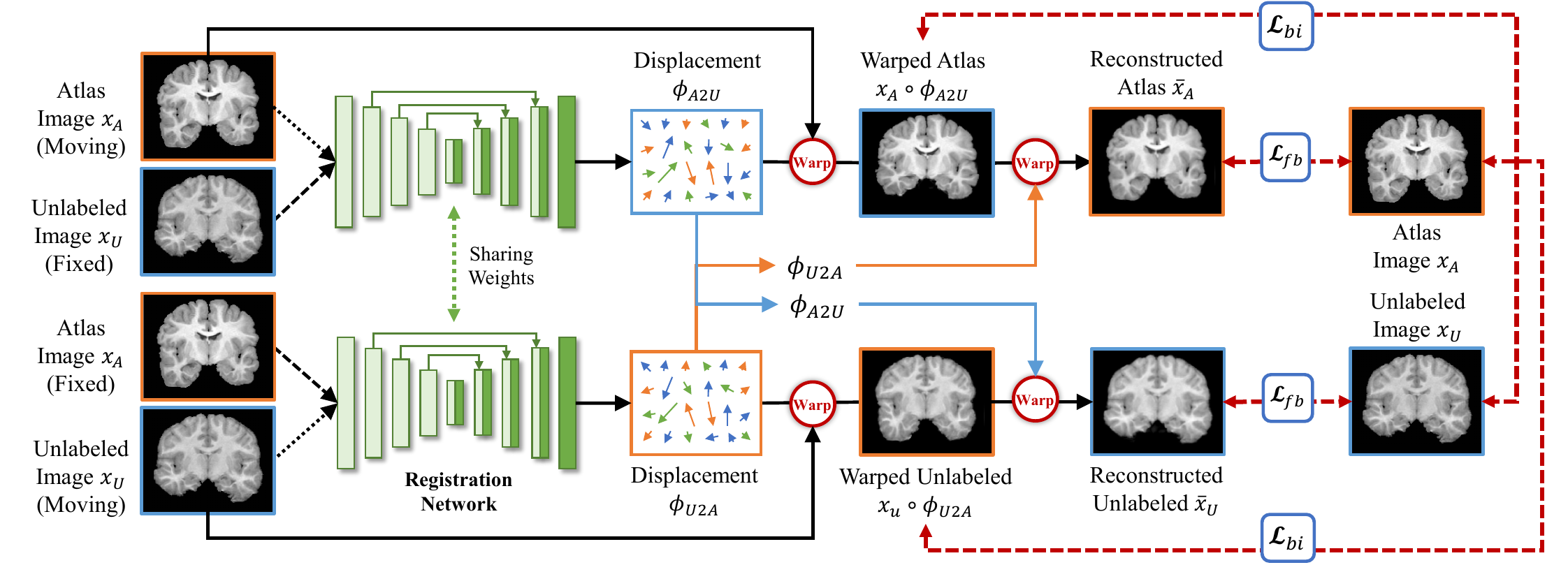}
    \caption{The overview of proposed anatomy-aligned registration with dual consistency constraint. Our dual consistency constraint encourages a reversible and bidirectional registration. First, the warped atlas/fixed images are deformed back to the original topology to ensure that topology alignment in registration is reversible. Then, the roles of moving/fixed images are inversed during registration to guarantee the bidirectional consistency.}
    \label{fig:reg}
\end{figure*}

\section{Method}
\label{sec:method}
In this section, we introduce our AdLER method as an universal one-shot medical image segmentation framework. We commence by introducing the preliminaries and the basic workflow of AdLER, and then provide detailed descriptions of the anatomy-aligned registration, adversarial data augmentation, and label error rectification, respectively. 
Finally, we provide the pseudo-code to summarize the training of our AdLER method.

\subsection{Overview}
\subsubsection{Preliminaries on OSSLT}
Formally, given a single labeled image $x_A$ and its label $y_A$ as the atlas, the training objective of OSSLT is to learn segmentation $\mathcal{S}$ from the atlas $\mathcal{A}=(x_A, y_A)$ and a large unlabeled dataset $\mathcal{D}_{U}=\{x_{U1},x_{U2},...\}$ based on deformable registration $\mathcal{R}$. OSSLT generally consists of three steps: First, deformable registration $\mathcal{R}$ is learned from the unlabeled set $\mathcal{D}_{U}$ in an unsupervised manner. This step is usually achieved by widely recognized registration methods such as VoxelMorph \cite{balakrishnan2019voxelmorph} and \etc \cite{wang2020lt, he2022learning}. Next, the spatial and appearance reference images (denoted as $x_s$ and $x_a$) are sampled from the unlabeled set $\mathcal{D}_{U}$ and fed to the registration network $\mathcal{R}$. After that, the corresponding spatial transformation $\phi \in \mathbb{R}^3$ and appearance transformation $\psi \in \mathbb{R}$ are generated to perform data augmentation on the atlas $\mathcal{A}$. For the spatial transformation, the atlas image $x_A$ and the spatial reference $x_s$ are fed to the registration network $\mathcal{R}$ to predict a deformation field $\phi$, which is used as the spatial transformation. For appearance transform, the atlas image $x_A$ and the appearance reference $x_a$ are fed to $\mathcal{R}$ as well. Unlike the spatial transformation, the registration here is the inverse registration $\phi^{-1}_{xa}$, which warps $x_a$ to $x_A$ and generates an inverse-warped appearance reference $\tilde{x_a}=x_a \circ \phi^{-1}_{xa}$, and appearance transformation $\psi=\tilde{x_a} - x_A$ is generated by the residual of inverse-warped appearance reference $\tilde{x_a}$ and the atlas image $x_A$.

During training, a large pseudo-labeled set $\mathcal{D}_{PL}=\{(x_{g1},y_{g1}),(x_{g2},y_{g2}),...\}$ is generated by sampling different reference images from the unlabeled set $\mathcal{D}_{U}$, and augmenting the atlas $\mathcal{A}$ with the transformations. Specifically, given the spatial transformation $\phi$ and the appearance transformation $\psi$, the augmented image $x_g=(x_A+\psi) \circ \phi$ is generated by applying both transformations, and the warped label $y_g=y_A \circ \phi$ serves as the ground truth. The segmentation network $\mathcal{S}$ is learned from the pseudo-labeled set $\mathcal{D}_{PL}$.


\subsubsection{Overall Pipeline of AdLER}
Our one-shot medical image segmentation framework consists of three networks, which are the registration network $\mathcal{R}$, the adversarial network $\mathcal{G}$ and the segmentation network $\mathcal{S}$. The entire framework with the three networks is constructed in an end-to-end manner, and the basic diagram of our one-shot medical segmentation method is illustrated in Fig. \ref{fig:OSSLT} (b). Note that in the training stage, we commence by optimizing the registration network as it is needed to learn the data distribution of the unlabeled data. Then, we freeze the registration $\mathcal{R}$ and train the augmentation $\mathcal{G}$ and segmentation $\mathcal{S}$ in an adversarial manner. Following \cite{zhao2019data, ding2021modeling, he2022learning}, two unlabeled images are sampled from the training set as spatial and appearance references to generate the corresponding transformation to augment the single atlas image. After acquiring the augmented image $x_g$, it is then fed to the adversarial network to generate two adversarial transformations, which can improve both the diversity of the transformations and the robustness of segmentation against perturbations. The adversarial training here is the "min-max" game between two networks \cite{goodfellow2014explaining}. Specifically, given a segmentation $\mathcal{S}$, the adversarial transformation generated by adversarial network $\mathcal{G}$ is expected to maximize the segmentation difference before and after applying the adversarial transformation. The adversarial augmented images are then fed to the segmentation network $\mathcal{S}$ to train the segmentation model, where the learning objective is to minimize the segmentation difference mentioned above, as well as the common segmentation losses.

\subsection{Anatomy-Aligned Registration}
\label{sec:registration}
To ensure the effectiveness of adversarial training and rectified segmentation, we introduce an anatomy-aligned registration to improve anatomical alignment and topology preservation. To this end, we propose a dual consistency constraint in registration as shown in Fig. \ref{fig:reg}. The dual consistency constraint is developed to encourage registration to be bidirectional and reversible. Note that unlike CycleMorph \cite{kim2021cyclemorph} which achieves similar constraints by two networks and multiple forward runs, our dual consistency constraint only requires one single registration network.

Specifically, given the atlas image $x_A$ and the unlabeled image $x_U$, we also introduce the bidirectional consistency in topology during registration to ensure its effectiveness:

\begin{equation}
    \phi_{A2U}=\mathcal{R}(x_A, x_U), \ \phi_{U2A}=\mathcal{R}(x_U, x_A)
\end{equation}
where $\phi_{A2U}$ and $\phi_{U2A}$ denote the deformation fields that align $x_A$ to $x_U$ and $x_U$ to $x_A$, respectively.

Theoretically, if an ideal registration can perfectly align the topology of $x_A$ to $x_U$, then the inverse registration that aligns $x_U$ to $x_A$ should also align the warped $x_A$ back to $x_A$, and vice versa. In other words, the registration should be reversible. Thus, we propose to reuse the deformation fields $\phi_{A2U}$ and $\phi_{U2A}$ calculated above to regularize the consistency of forward registration and backward reconstruction:
\begin{equation}
    \bar{x}_{A}=(x_A \circ \phi_{A2U}) \circ \phi_{U2A}, \ \bar{x}_{U}=(x_U \circ \phi_{U2A}) \circ \phi_{A2U},
\end{equation}
where $\bar{x}_{A}$ and $\bar{x}_{U}$ denote the reconstructed atlas image and unlabeled image, respectively.

The loss functions to optimize registration consist of two components, which are the bidirectional consistency $\mathcal{L}_{bi}$ and reversible consistency constraint $\mathcal{L}_{fb}$. For bidirectional consistency $\mathcal{L}_{bi}$, it consists of basic similarity loss $\mathcal{L}_{sim}$ and smooth regularization loss $\mathcal{L}_{smooth}$ to ensure bidirectional consistency in topology for the registration. We also adopt the L2 loss as the similarity loss $\mathcal{L}_{sim}$, and bending energy loss \cite{rueckert1999nonrigid} as the smooth regularization loss $\mathcal{L}_{smooth}$. The bidirectional consistency $\mathcal{L}_{bi}$ is written as follows:
\begin{equation}
\begin{aligned}
    & \mathcal{L}_{bi}(x_A,x_U,\phi_{A2U},\phi_{U2A}) \\
    &= \mathcal{L}_{sim}(x_A \circ \phi_{A2U}, x_U) + \mathcal{L}_{sim}(x_U \circ \phi_{U2A}, x_A) \\
    &+ \lambda_{smooth}(\mathcal{L}_{smooth}(\phi_{A2U}) + \mathcal{L}_{smooth}(\phi_{U2A}))
\end{aligned}
\end{equation}

For the reversible consistency constraint $\mathcal{L}_{fb}$, we encourage both image-level consistency and semantic-level consistency for the original and reconstructed images. For image-level consistency, L1 loss is utilized to ensure the structural similarity between the original and the reconstructed images. For semantic-level consistency, both images are fed to the segmentation network $\mathcal{S}$ to encourage consistency in their segmentation results. The reversible consistency constraint $\mathcal{L}_{fb}$ is written as: 
\begin{equation}
\begin{aligned}
    & \mathcal{L}_{fb}(\bar{x}_{A}, \bar{x}_{U}, x_A, x_U) \\
    &= \mathcal{L}_{L1}(\bar{x}_{A}, x_A) + \mathcal{L}_{L1}(\bar{x}_{U}, x_U) \\
    &+ \lambda_{Dice}[\mathcal{L}_{Dice}(\mathcal{S}(\bar{x}_{A}), \mathcal{S}(x_A)) + \mathcal{L}_{Dice}(\mathcal{S}(\bar{x}_{U}), \mathcal{S}(x_U))]
\end{aligned}
\end{equation}

Finally, the overall registration loss $\mathcal{L}_{reg}=\mathcal{L}_{bi}(x_A,x_U,\phi_{A2U},\phi_{U2A})+\mathcal{L}_{fb}(\bar{x}_{A}, \bar{x}_{U}, x_A, x_U)$ is the linear combination of bidirectional consistency and reversible consistency constraint.

\subsection{Adversarial Training for Data Augmentation}
\begin{figure*}[t]
    \centering
    \includegraphics[width=0.90\textwidth]{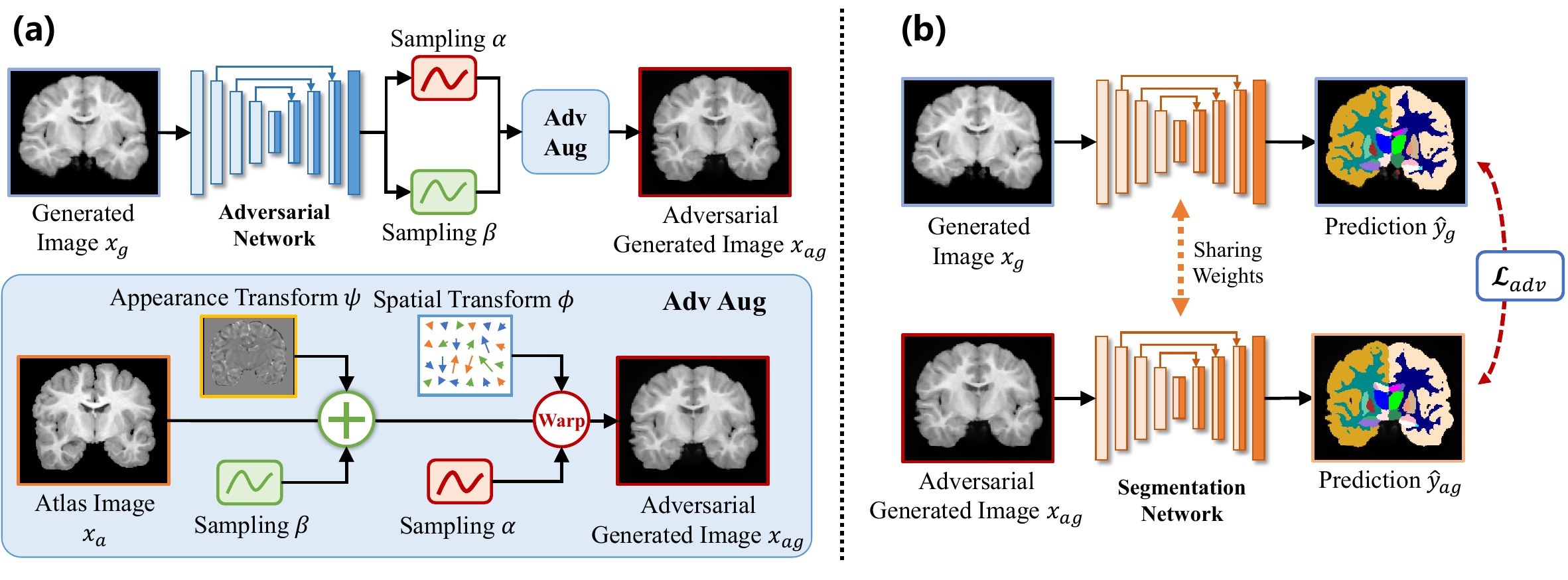}
    \caption{The overall workflow of adversarial training: (a) The vanilla augmented image $x_g$ is further augmented in an adversarial manner to generate an adversarial augmented image $x_{ag}$; (b) The segmentation network is trained to minimize the segmentation difference before and after adversarial augmentation.}
    \label{fig:adversarial}
\end{figure*}

The diversity of the pseudo-labeled set $\mathcal{D}_{PL}$ has a substantial impact on the segmentation performance. Therefore, it is essential to further enhance the diversity of transformations derived from registration. We propose to introduce adversarial training to transformation sampling to improve the diversity of the pseudo-labeled set $\mathcal{D}_{PL}$. The transformations used to augment the atlas are additionally influenced by sampling layers to add additional variance to the augmented data, while the segmentation is trained to learn under adversarial perturbations.
The framework of adversarial training is shown in Fig. \ref{fig:adversarial}.

Similar to common OSSLT methods, spatial transformation $\phi$ and appearance transformation $\psi$ are extracted from corresponding spatial and appearance reference images. Given the transformation $\phi$ and $\psi$, the augmented image $x_g=(x_A+\psi) \circ \phi$ is obtained by applying both transformations. However, this augmentation protocol is highly dependent on variant reference images to ensure generation diversity. Thus, we propose to introduce an adversarial network $\mathcal{G}$ to learn the transformation perturbations in an end-to-end manner. Specifically, the vanilla augmented image $x_g$ is fed to the adversarial network $\mathcal{G}$ to generated two sampling layers $\alpha$ and $\beta$: 
\begin{equation}
    \alpha, \beta = \mathcal{G}(x_g)
\end{equation}
where $\alpha \in \mathbb{R}^3$ is activated by the Sigmoid function and $\beta \in \mathbb{R}$ is activated by the Tanh function.

The sampling layers are combined with the original transformations $\phi$ and $\psi$ to add extra perturbations, which improves the diversity of the generated transformations.
\begin{equation}
    \phi_{a}=\alpha \times \phi, \ \psi_{a}=\psi+\beta \times \bar{\psi}
\end{equation}
where $\bar{\psi}$ is the mean value of $\psi$.

Technically, spatial sampling $\alpha$ and appearance sampling $\beta$ represent the sampling amplitude of the original transformation, ranging from 0 to 1 and -1 to 1, respectively. The spatial transformation sampling is used to simulate the spatial variance of medical images in the population, and the appearance transformation sampling is designed to follow the inhomogeneity in medical images. Thus, by applying the adversarial transformation $\phi_{a}$ and $\psi_{a}$ to the atlas image $x_A$, we obtain an adversarial augmented sample $(x_{ag},y_{ag})$:
\begin{equation}
    x_{ag}=(x_A+\psi_a) \circ \phi_a, \ y_{ag}=y_A \circ \phi_a
\end{equation}

Both original augmented image $x_g$ and adversarial augmented image $x_{ag}$ are fed to the segmentation network, and the training objective is the min-max game of the adversarial network $\mathcal{G}$ and the segmentation network $\mathcal{S}$. Thus, we ensure the diversity of generation and robustness of segmentation simultaneously by the adversarial training strategy written as:
\begin{equation}
    \min_{\mathcal{S}} \max_{\mathcal{G}} \mathcal{L}_{adv}(\hat{y}_{g}, \hat{y}_{ag})
\end{equation}
\begin{equation}
    \mathcal{L}_{adv}(\hat{y}_{g}, \hat{y}_{ag})=\frac{{\hat{y}_{g}} \cdot \hat{y}_{ag}}{\parallel \hat{y}_{g}\parallel_2 \cdot \parallel \hat{y}_{ag}\parallel_2}
\end{equation}
where $\hat{y}_{g}$ and $\hat{y}_{ag}$ denote the segmentation predictions of $x_g$ and $x_{ag}$, respectively. Since the spatial transformations applied to $x_g$ and $x_{ag}$ are different, the loss calculation is performed in the atlas space by inverse registration.
\begin{figure}[h]
    \centering
    \includegraphics[width=0.45\textwidth]{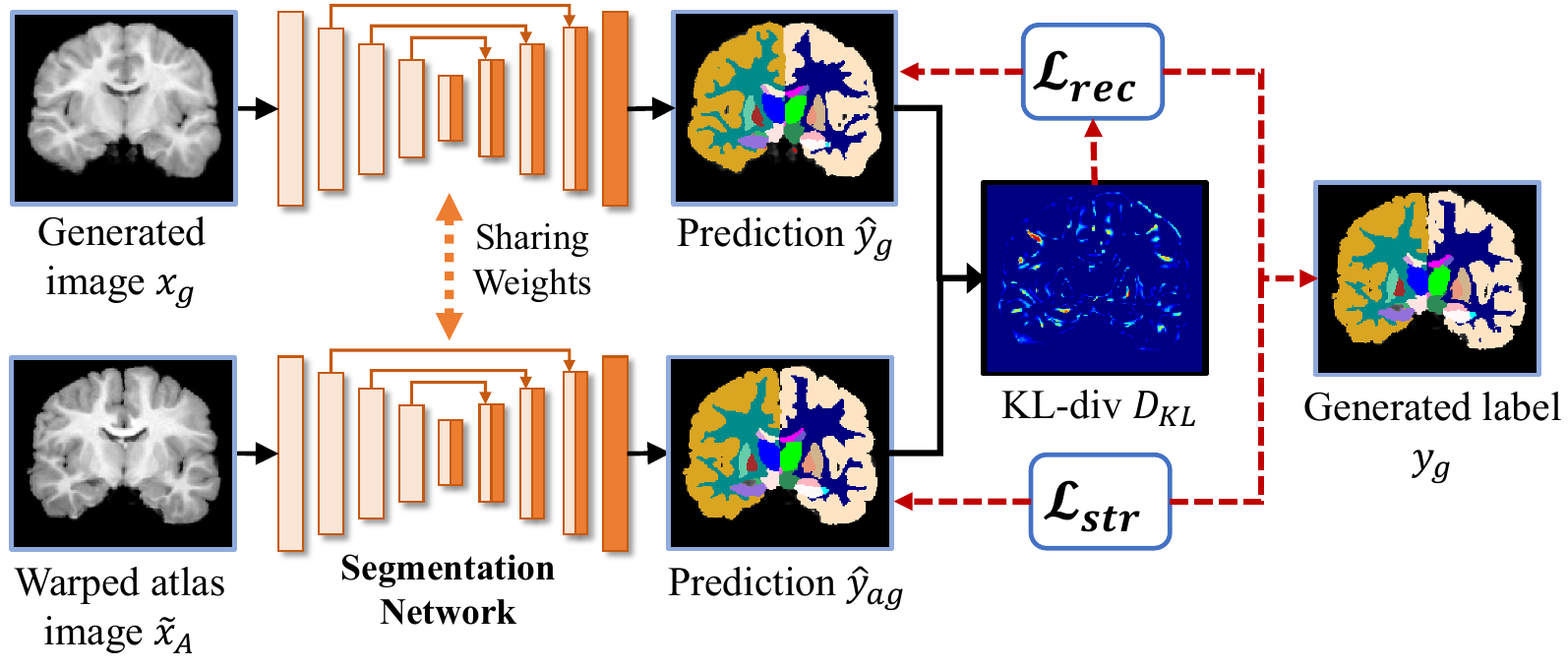}
    \caption{Illustration of uncertainty-based label error rectification proposed in AdLER. We use the KL-divergence of the predictions to estimate the potential label errors and surpass them.}
    \label{fig:uncertainty}
\end{figure}

\subsection{Label Error Rectification for Segmentation}
Most of the current OSSLT methods hypothesize that the appearance transformation does not affect the authenticity of the label. However, this is not true due to the imperfect nature of registration. Inspired by \cite{zheng2021rectifying}, we propose to rectify the label errors based on uncertainty estimation, which is shown in Fig. \ref{fig:uncertainty}. To rectify the label errors introduced by the appearance transformations, two augmented inputs, augmented image $x_g=(x_A+\psi) \circ \phi$ with both spatial and appearance transformations, and warped atlas $\tilde{x}_A=x_A \circ \phi$ augmented by only spatial transformation, are fed into the segmentation network $\mathcal{S}$, respectively. The main idea is to utilize the spatial and appearance transformations separately: The warped atlas $\tilde{x}_A$ has more label authenticity, which is used to guide the segmentation to learn the simulated spatial variance of medical images. On the other hand, compared with warped atlas $\tilde{x}_A$, augmented image $x_g$ is equipped with more diversity by combining both spatial and appearance transformations, and guides the segmentation to learn segmentation under various augmentation. 

The supervised segmentation loss $\mathcal{L}_{seg}$ consists of two items, which are the structure loss $\mathcal{L}_{str}$ and the rectification loss $\mathcal{L}_{rec}$: For the warped atlas $\tilde{x}_A$, the network is expected to learn the spatial variance of medical images; For the augmented image $x_g$, the network should learn segmentation under both transformations, despite the potential label errors introduced by appearance transformation. In this way, the loss functions are described as follows:
\begin{equation}
    \mathcal{L}_{str}(\hat{\tilde{y}}_A, y_g)=\mathcal{L}_{Dice}(\hat{\tilde{y}}_A, y_g) + \mathcal{L}_{ce}(\hat{\tilde{y}}_A, y_g)
\end{equation}
\begin{equation}
\begin{aligned}
    \mathcal{L}_{rec}(\hat{y}_g, \hat{\tilde{y}}_A, y_g)&=\exp[-D_{KL}(\hat{y}_g, \hat{\tilde{y}}_A)]\mathcal{L}_{ce}(\hat{y}_g, y_g) \\ 
    &+ \lambda_{KL} D_{KL}(\hat{y}_g, \hat{\tilde{y}}_A)
\end{aligned}
\end{equation}
where $\hat{y}_g$ and $\hat{\tilde{y}}_A$ denote the predictions of $x_g$ and $\tilde{x}_A$, respectively.

For structure loss $\mathcal{L}_{str}$, we use the combination of Dice loss $\mathcal{L}_{Dice}$ \cite{milletari2016v} and cross-entropy loss $\mathcal{L}_{ce}$, which are widely used in medical image segmentation. For the rectification loss $\mathcal{L}_{rec}$, we follow \cite{zheng2021rectifying} and adopt the KL-divergence $D_{KL} $ of the segmentation results $\hat{\tilde{y}}_A$ and $\hat{y}_g$ as uncertainty in the prediction. As the voxel with higher uncertainty indicates greater possibility of label error at the corresponding location of $x_g$, its corresponding supervision signal should be weakened to reduce the effect of label errors. Thus, the exponential KL-divergence serves as the weight of the cross-entropy loss at every location. It should be noted that minimizing the weighted cross-entropy loss $\exp[-D_{KL}(\hat{y}_g, \hat{\tilde{y}}_A)]\mathcal{L}_{ce}(\hat{y}_g, y_g)$ directly can simply improve the uncertainty of the network predictions. Thus, KL-divergence is employed as a regularization term to regularize the training.

Thus, the overall supervised segmentation loss $\mathcal{L}_{seg}=\mathcal{L}_{str}(\hat{\tilde{y}}_A, y_g)+\lambda_{rec}\mathcal{L}_{rec}(\hat{y}_g, \hat{\tilde{y}}_A, y_g)$ is the combination of structure loss and rectification loss. We apply the supervised segmentation loss $\mathcal{L}_{seg}$ in both the vanilla augmented image $x_g$ and adversarial augmented image $x_{ag}$ in practice.

In summary, the whole process for training our AdLER is presented in Algorithm \ref{sec:algorithm}.

\begin{algorithm}[h]
\caption{Pseudo-code for training AdLER}
\label{sec:algorithm}

\KwIn{Registration network $\mathcal{R}$; Augmentation network $\mathcal{G}$; Segmentation network $\mathcal{S}$}
\KwOut{Trained segmentation network $\mathcal{S}$ for inference}
\KwData{Atlas $(x_A,y_A)$; Unlabeled dataset $\mathcal{D}_{U}$}
Set the target training iteration number $N_{all}$\;

\For{i {\rm in} range($N_{all}$)}{
Sample reference images $x_s$ and $x_a$ from $\mathcal{D}_{U}$\;

\tcp{first train registration}
\While{$\mathcal{R}$.{\rm requires\_grad}=\textit{True}}{
Train the registration $\mathcal{R}$ and get the loss $\mathcal{L}_{reg}$\;
Optimize the registration: $\min\limits_{\mathcal{R}} \mathcal{L}_{reg}$ \;
}

Freeze the registration: $\mathcal{R}$.requires\_grad=\textit{False}\;
Generate transformations $\phi$ and $\psi$\;
Augment the atlas: $x_g=(x_A+\psi) \circ \phi$;$y_g=y_A \circ \phi$\;

\tcp{train augmentation}
\While{$\mathcal{G}$.{\rm requires\_grad}=\textit{True}}{
Generate transformation sampling $\alpha$ and $\beta$\;
Adversarial augmentation and get  $x_{ag}$\;
Segment both $x_g$ and $x_{ag}$\;
Optimize the augmentation: $\max\limits_{\mathcal{G}} \mathcal{L}_{adv}$\;
}
Freeze the augmentation: $\mathcal{G}$.requires\_grad=\textit{False}\;

\tcp{train segmentation}
\While{$\mathcal{S}$.{\rm requires\_grad}=\textit{True}}{
Generate transformation sampling $\alpha'$ and $\beta'$\;
Adversarial augmentation and get  $x_{ag}'$\;
Segment both $x_g$ and $x_{ag}'$\;
Rectified segmentation: $\mathcal{L}_{seg}=\mathcal{L}_{str}+\mathcal{L}_{rec}$\;
Optimize the segmentation: $\min\limits_{\mathcal{S}} (\mathcal{L}_{adv}+\mathcal{L}_{seg}$)\;
}
Freeze the segmentation: $\mathcal{S}$.requires\_grad=\textit{False}\;
Unfreeze the registration: $\mathcal{R}$.requires\_grad=\textit{True}\;
}
\end{algorithm}

\section{Experiments}
\label{sec:experiments}
\subsection{Datasets}
In this section, we introduce the two public datasets used to validate the effectiveness of the proposed AdLER, which are the CANDI \cite{kennedy2012candishare} and ABIDE\cite{di2014autism} datasets.
\begin{table*}[]
\centering
\caption{Segmentation DSC (\%) of different algorithms on CANDI and ABIDE datasets. The best results are highlighted in bold.}
\label{table:compare}
\renewcommand\arraystretch{1.05}
\setlength{\tabcolsep}{2.5mm}{
\begin{tabular}{ccccclcccc}
\toprule
\multirow{3}{*}{\textbf{Method}} & \multicolumn{3}{c}{\multirow{2}{*}{\textbf{CANDI}}} & \multicolumn{6}{c}{\textbf{ABIDE}}                                                                              \\ \cline{5-10} 
                                              & \multicolumn{3}{c}{}                                & \multicolumn{3}{c}{\textbf{Seen}}                                    & \multicolumn{3}{c}{\textbf{Unseen}}                        \\ \cline{2-10}
                                     & Mean ± Std        & Min         & Max        & Mean ± Std & \multicolumn{1}{c}{Min} & Max  & Mean ± Std   & Min   & Max   \\ 
\midrule
DataAug \cite{zhao2019data}          & 80.4 ± 4.3        & 73.8        & 84.0       & 69.6 ± 9.0 & 33.1                    & 82.5 & 64.3 ± 9.9   & 32.3  & 79.6  \\
LT-Net \cite{wang2020lt}             & 82.3 ± 2.5        & 75.6        & 84.2       & 71.3 ± 9.2 & 39.7                    & 80.4 & 66.1 ± 11.4  & 35.0  & 77.2  \\
VAEAug \cite{ding2021modeling}       & 85.1 ± 1.9        & 80.2        & 87.8       & 76.7 ± 7.4 & 53.2                    & 86.5 & 74.8 ± 6.6   & 54.1  & 83.3  \\
BRBS \cite{he2022learning}           & 85.7 ± 1.0        & 81.3        & 87.5       & 77.5 ± 8.6 & 54.4                    & 87.6 & 74.5 ± 7.1   & 53.8  & 83.6  \\
\midrule
\textbf{AdLER}                       & \textbf{86.4 ± 1.5}  & \textbf{83.3}  & \textbf{88.6} & \textbf{81.1 ± 7.9} & \textbf{55.2}           & \textbf{89.5} & \textbf{79.4 ± 3.3} & \textbf{70.4} & \textbf{86.1} \\
\bottomrule
\end{tabular}
}
\end{table*}

\subsubsection{CANDI Dataset}
The CANDI dataset \cite{kennedy2012candishare} contains 103 brain T1 MR images from 57 males and 46 females, and includes 28 primary brain regions for manual annotations. There are four types of diagnostic groups included in this dataset, which are healthy controls, schizophrenia spectrum, bipolar disorder with psychosis, and bipolar disorder without psychosis. Note that we use the same data split, label split and preprocessing as previous works \cite{wang2020lt, ding2021modeling, he2022learning} for fair comparison: We implement the data split as 1, 82, 20 images as the atlas, unlabeled training set and the test set, respectively. Note that we standardize the image size to 160$\times$160$\times$128  by cropping around the center of the brain region during training and inference, ensuring sufficient coverage of the entire brain area. 

\subsubsection{ABIDE Dataset}
The ABIDE dataset \cite{di2014autism} is collected from 17 international imaging sources. Different with CANDI dataset \cite{kennedy2012candishare}, ABIDE dataset is considered more appropriate to assess the performance of the proposed AdLER in the clinical scenario, where it is expected to exhibit robust performance across different image sources. Note that Ding \etal \cite{ding2021modeling} has established a benchmark for the one-shot segmentation task in the ABIDE dataset, and we use the same setting to evaluate our proposed AdLER. Specifically, the ABIDE dataset contains two tasks, where the one-shot segmentation method is evaluated on the "seen" and "unseen" data sites, respectively: First, 100 images from 10 imaging sources are sampled as the training dataset; For the test dataset, 60 images from the same imaging sources and 60 images from the rest 7 imaging sources are selected as the "seen" and "unseen" test sets, respectively. As suggested in \cite{balakrishnan2019voxelmorph}, we choose one image that is closest to the average volume of the training set as the atlas. All images are resampled into 1mm$^3$ isotropic voxels, which are then centerly cropped to unify the image size as 160$\times$160$\times$192. Note that the same 28 brain ROIs following the annotations in CANDI dataset are employed in the ABIDE dataset to evaluate the segmentation performance of AdLER.

\subsection{Implementation Details}
The proposed AdLER is implemented with PyTorch 1.12.1 platform on a Debian server. Distributed data-parallel configurations with 2 NVIDIA RTX 3090 GPUs are applied to accelerate the training process. PyTorch native automatic mixed-precision training is employed to reduce video memory cost. MONAI library \cite{monai} has been used to accelerate the data loading process. For the network backbone, we adopt 3D U-Net \cite{ronneberger2015u} to construct the registration, augmentation and segmentation networks, which are optimized by SGD optimizer with a weight decay of $1 \times 10^{-5}$. The initial learning rate is set to $1 \times 10^{-2}$ after a warm-up for 5 epochs, and is slowly reduced with the cosine annealing strategy. The network is trained for $5 \times 10^{4}$ iterations in total and initiates early stop if performance starts to plateau. The batch size is set to 1 due to limits in GPU memory resources. For hyperparameter settings, we set $\lambda_{smooth}=15.0$ and $\lambda_{Dice}=10.0$ in registration following \cite{he2022learning}, while $\lambda_{KL}=1 \times 10^{-4}$ and $\lambda_{rec}=0.5$ in segmentation. The weight of adversarial loss and supervised segmentation loss are set as the same when training segmentation. It should be noted that since the early stage of segmentation is inaccurate, we set $\lambda_{Dice}$, $\lambda_{KL}$ and $\lambda_{rec}$ as a time-dependent Gaussian warming-up function \cite{luo2022semi, wu2022mutual} to stabilize the training. 

For performance comparison, we compare the proposed AdLER with previous cutting-edge one-shot medical image segmentation methods, including DataAug \cite{zhao2019data}, LT-Net \cite{wang2020lt}, VAEAug \cite{ding2021modeling} and BRBS \cite{he2022learning}. We employ the Dice similarity coefficient (DSC) to evaluate the performance of different one-shot segmentation methods.
\begin{figure*}[h]
	\centering
	\includegraphics[width=0.90\textwidth]{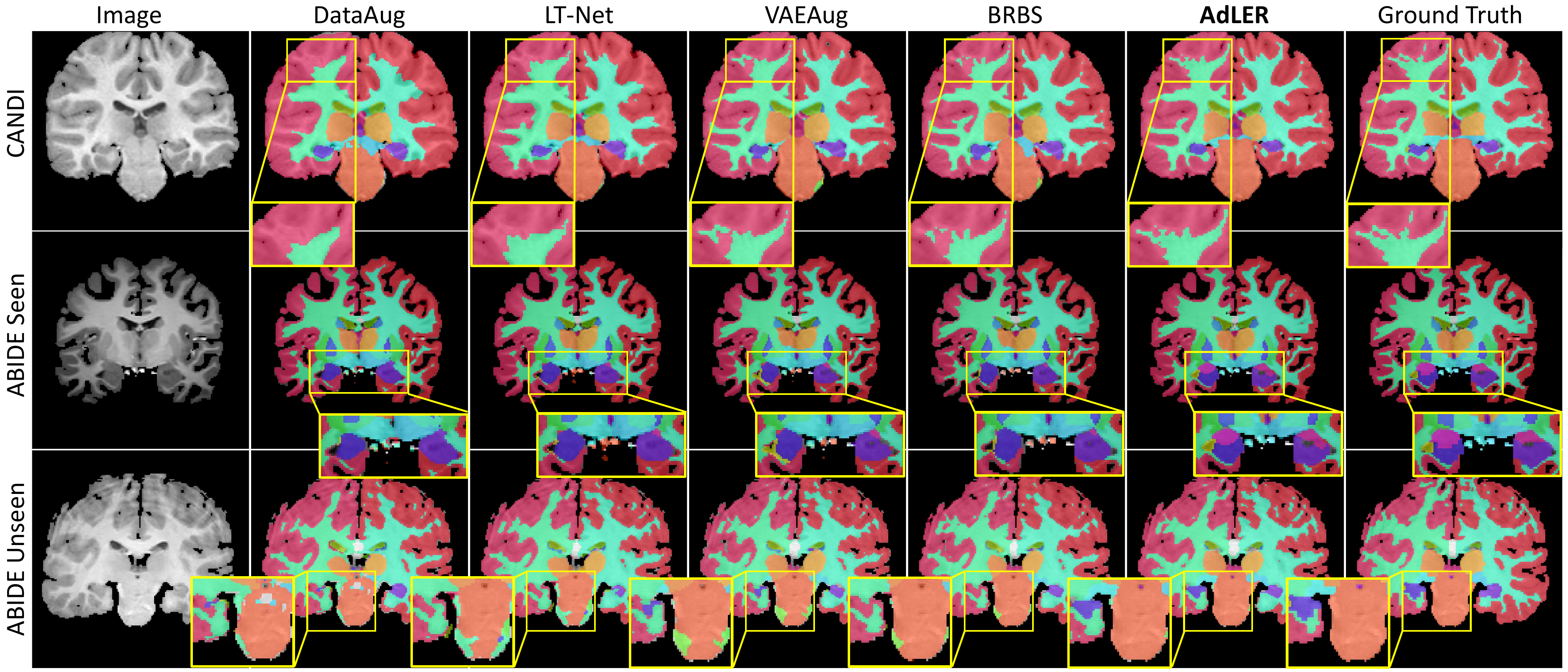} 
	\caption{Visual comparison of different one-shot medical image segmentation methods on two public datasets.}
	\label{fig:comparison}
\end{figure*}

\subsection{Results}
\subsubsection{Quantitative Results}
Comparisons of the proposed AdLER with cutting-edge alternatives in the two public-available datasets are shown in Table \ref{table:compare}. In the CANDI dataset, AdLER yields an average Dice similarity coefficient of 86.4\% which surpasses all previous works, and also achieves the best score in both the minimum and maximum segmentation performance. Furthermore, AdLER has the superiority in its robustness that it also significantly improves the segmentation performance of the worst case as observed in Table \ref{table:compare}. 

For the ABIDE dataset, our AdLER also achieves superior performance in both the "seen" and "unseen" test sets, which exceeds the state-of-the-art methods by 3.6\% and 4.9\% in the "seen" and "unseen" scenarios, respectively. The experimental results on the ABIDE dataset prove the effectiveness of the proposed AdLER on multi-center data, demonstrating the potential of our method in real-world scenarios. Also note that the proposed AdLER has better average segmentation performance with a relatively lower standard deviation, which also reflects its effectiveness and robustness in OSSLT. It can be observed that AdLER obtains a greater performance leap in the "unseen" dataset as well, with a much higher average performance and a lower standard deviation.

\subsubsection{Qualitative Results}
We have visualized the segmentation results of different one-shot medical segmentation methods on the two public datasets, which is shown in Fig. \ref{fig:comparison}. The yellow bounding boxes highlight the regions where our AdLER outperforms the alternatives. Under adversarial training, AdLER yields better segmentation robustness and achieves better segmentation performance, especially for the regions where brain topologies are complicated. For instance, in the second and third rows, other one-shot segmentation methods are hard to differentiate the trivial details in the anatomical structures of the brain, while our method can segment them precisely.
Furthermore, with the integration of the proposed label error rectification mechanism, our AdLER has improved performance compared to the alternatives when delineating tissue boundary regions, as shown in the first row of Fig. \ref{fig:comparison}. 

\subsection{Ablation Studies}
Here we conduct ablation studies to validate the effectiveness of the proposed modules in AdLER. We commence by evaluating the effectiveness of the proposed anatomy-aligned registration method in AdLER, which is compared with the registration used in previous OSSLT algorithms. Then, we investigate the adversarial training in augmentation and label error rectification in segmentation that are proposed in the AdLER method.
\subsubsection{Effectiveness of Anatomy-Aligned Registration}
We conduct experiments on the CANDI dataset to compare the proposed registration method with the alternatives, which is shown in Table \ref{table:registration}. Note that adversarial training and label error rectification are excluded to avoid the potential effect of these proposed modules on registration. It is shown in Table \ref{table:registration} that the proposed registration method in our AdLER yields the best performance in both registration and segmentation, which demonstrates the advantage of our registration method compared to VoxelMorph used in \cite{zhao2019data}. Furthermore, our registration method outperforms the registration utilized in the works of \cite{wang2020lt} and \cite{he2022learning}, demonstrating the superiority of our registration method over the state-of-the-art OSSLT methods.
\begin{table}[]
\centering
\caption{Comparison of registration methods on CANDI dataset. Note that we use common adversarial training and label error rectified segmentation in this experiment.}
\label{table:registration}
\begin{tabular}{ccc}
\toprule
\textbf{Registration} & \textbf{Reg. DSC (\%) $\uparrow$}   & \textbf{Seg. DSC (\%) $\uparrow$} \\
\midrule
VoxelMorph \cite{balakrishnan2019voxelmorph}   & 75.4 ± 9.2 & 80.4 ± 4.3   \\
LT-Net \cite{wang2020lt}       & 72.9 ± 7.5 & 78.9 ± 5.8   \\
BRBS \cite{he2022learning}         & 76.6 ± 6.2 & 80.5 ± 4.9   \\
\midrule
AdLER     & \textbf{77.8 ± 5.1} & \textbf{81.1 ± 4.0} \\
\bottomrule
\end{tabular}
\end{table}

\subsubsection{Effectiveness of Adversarial Training and Label Error Rectification}
Further experiments have been conducted to validate the proposed adversarial training and label error rectification. Quantitative results are shown in Table \ref{table:modules}. We begin by comparing the proposed adversarial training with the vanilla strategy in \cite{zhao2019data}, and the predefined transformation sampling based on Beta distribution proposed in \cite{he2022learning}. It can be observed in No.2 that transformation sampling based on Beta distribution brings slight performance improvement of 0.5\% compared to No. 1. Meanwhile, the adversarial training shown as No. 3 has achieved significant performance improvement of 1.8\% than No. 1, which demonstrates the effectiveness of our strategy. The segmentation performance has been further improved by 3\% in No.4 when combined with the proposed label error rectification mechanism. It can also be observed from No. 5 and No. 6 that the proposed label error rectification strategy can be combined with the augmentation techniques, and our AdLER which combines the adversarial training and label error rectification yields the best performance.
\begin{table}[]
\centering
\caption{Ablation studies on different augmentation and segmentation configurations on CANDI dataset with the proposed registration method. \textit{Beta} denotes transformation sampling from beta distribution, \textit{ADV} denotes adversarial training, and \textit{LER} denotes label error rectification.}
\label{table:modules}
\begin{tabular}{ccccc}
\toprule
\textbf{No.}         & \textbf{Beta}     & \textbf{ADV}     & \textbf{LER}    & \textbf{DSC \% $\uparrow$} \\ \midrule
1                    &          &         &        & 81.1 ± 4.0              \\
2                    & $\checkmark$        &         &        & 81.6 ± 3.4              \\
3                    &          & $\checkmark$       &        & 82.9 ± 2.7              \\
4                    &          &         & $\checkmark$      & 84.1 ± 2.9              \\
5                    & $\checkmark$        &         & $\checkmark$      & 84.5 ± 1.8              \\ \midrule
6 (AdLER)             &          & $\checkmark$       & $\checkmark$      & \textbf{86.4 ± 1.5}              \\ \bottomrule
\end{tabular}
\end{table}

\subsubsection{Visualization of Adversarial Augmented Images}
\begin{figure}[h]
	\centering
	\includegraphics[width=0.45\textwidth]{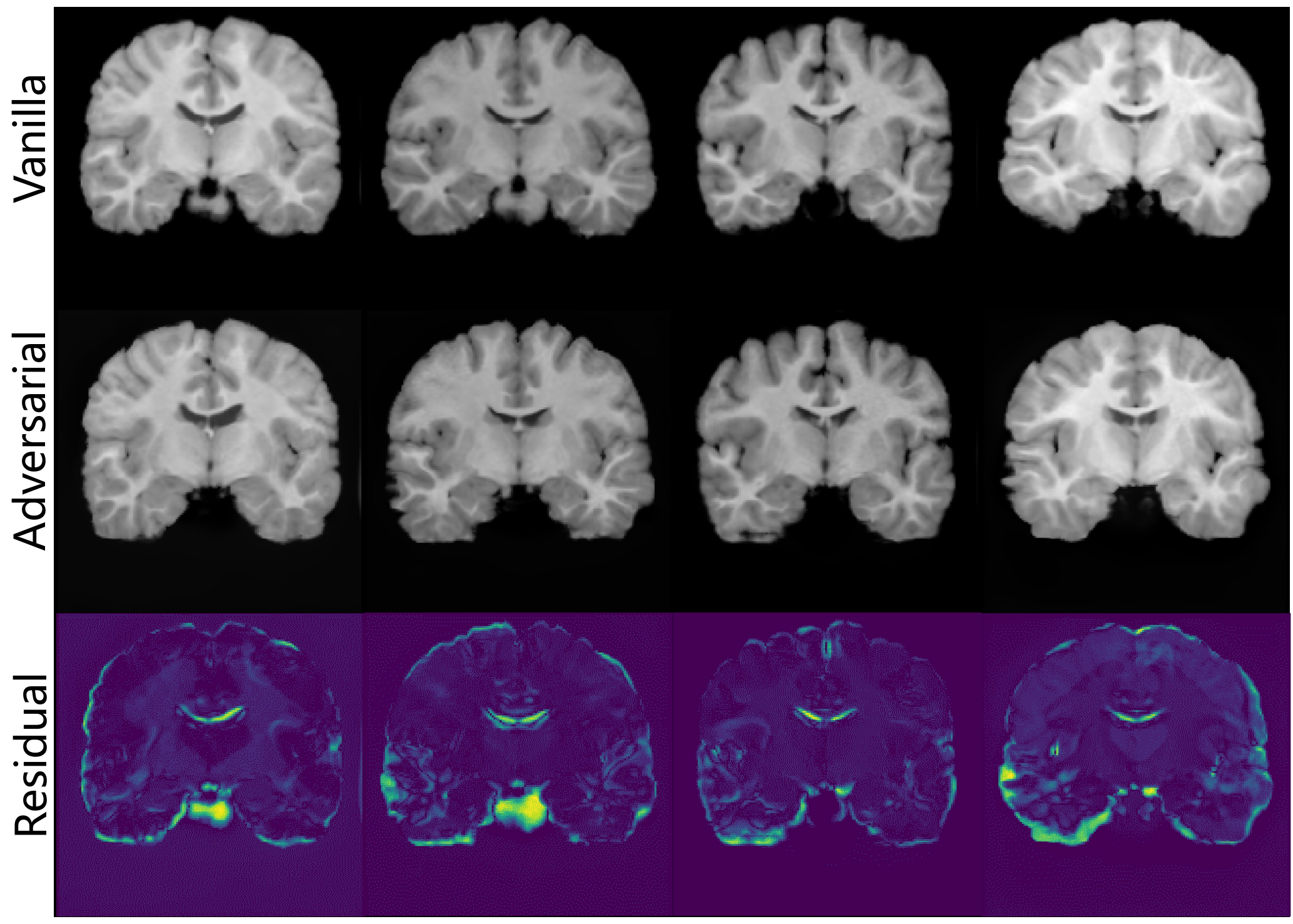} 
	\caption{Illustration of the augmented images and their corresponding adversarial augmented images. We also visualize their difference as residual images to illustrate how the image changes during adversarial augmentation.}
	\label{fig:generation}
\end{figure}

We have visualized several generated images before and after adversarial augmentation in Fig. \ref{fig:generation} to provide more intuitive illustrations of the proposed adversarial training strategy. The residual images presenting the absolute value of the subtraction in the two images are also shown in the third row to better observe their differences. It can be found that adversarial augmentation introduces more perturbations in the boundaries of adjacent brain tissues, where it is more challenging for segmentation methods to have accurate estimations. Besides, the intensity distribution of the generated images also varies after adversarial augmentation (see the white matter region in the first column), which can synthesize the inhomogeneity of MR scans across different centers, which is more appropriate for multi-center dataset such as ABIDE. 

\subsubsection{Adversarial Training for Data Scarcity}
\begin{figure}[h]
	\centering
	\includegraphics[width=0.45\textwidth]{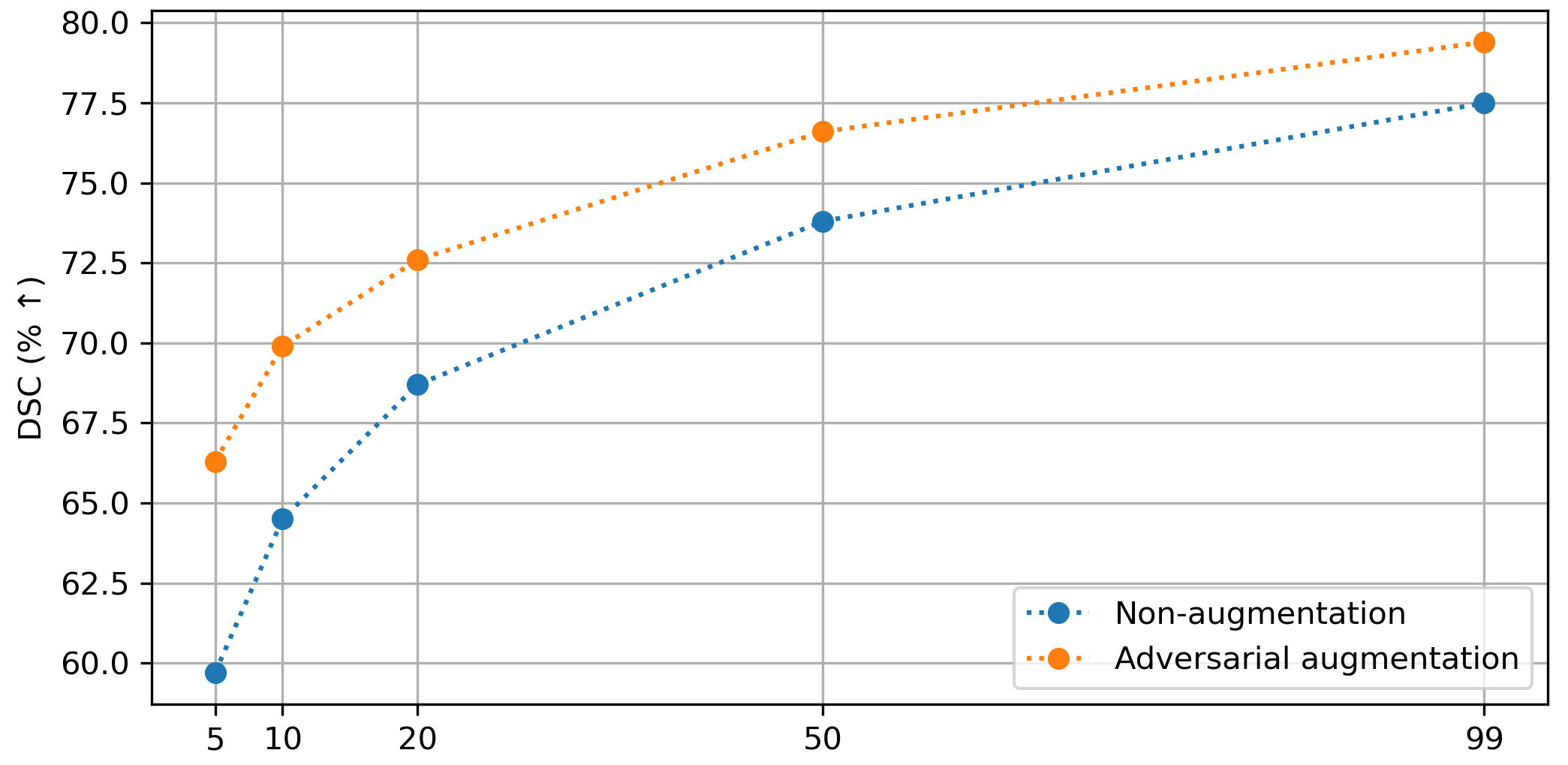} 
	\caption{Effects of adversarial training when faced with limited training samples on ABIDE "unseen" benchmark. The diagram presents the segmentation performance with or without adversarial augmentation in different numbers of unlabeled images.}
	\label{fig:adv_effect}
\end{figure}

Direct data augmentation based on reference images \cite{zhao2019data} sampled from the unlabeled dataset $\mathcal{D}_{U}$ greatly depends on the scale of $\mathcal{D}_{U}$. Limited availability of unlabeled images severely constrains the capacity for data augmentation, and the proposed adversarial training strategy can also alleviate such issues. We have conducted an intuitive experiment on the ABIDE "unseen" benchmark, where we intentionally reduce the number of unlabeled images and observe the trends of the performance change. The quantitative results are shown in Fig. \ref{fig:adv_effect}. As the number of unlabeled images decreases, both the segmentation performance under the non-augmentation setting in \cite{zhao2019data} and the proposed method decreases correspondingly. However, the performance drop of AdLER is significantly less pronounced, especially when the number of unlabeled images is severely limited. The results prove the success of the proposed adversarial training strategy when unlabeled data is limited. 

\section{Conclusion}
\label{sec:conclusion}
In this work, we have proposed a novel one-shot medical image segmentation method namely AdLER, to address the current challenging issues in the field of OSSLT. First, we develop an anatomy-aligned registration approach using a dual consistency constraint to provide registration with better structure alignment and alleviate label errors. Second, we propose an adversarial data augmentation pipeline for a min-max game between augmentation and segmentation, which can improve the diversity of the generated images and the robustness of segmentation. Furthermore, we introduce label error rectification based on the estimation of prediction uncertainty in the segmentation, to reduce the effect of label errors introduced during data augmentation. Experiments based on the CANDI and ABIDE datasets demonstrate the effectiveness of the proposed AdLER method, which can yield state-of-the-art segmentation performance compared with the alternatives. The proposed AdLER method is anticipated to have the potential for application in other one-shot and few-shot medical image segmentation tasks, which can further contribute to the developments of few-shot learning in the medical image segmentation domain.

\bibliography{references}

\begin{thebibliography}{10}

\bibitem{kaur2017survey}
G.~Kaur and J.~Chhaterji, ``A survey on medical image segmentation,'' {\em
  International Journal of Science and Research}, vol.~6, no.~4,
  pp.~1305--1311, 2017.

\bibitem{ronneberger2015u}
O.~Ronneberger, P.~Fischer, and T.~Brox, ``U-net: Convolutional networks for
  biomedical image segmentation,'' in {\em International Conference on Medical
  image computing and computer-assisted intervention}, pp.~234--241, Springer,
  2015.

\bibitem{isensee2021nnu}
F.~Isensee, P.~F. Jaeger, S.~A. Kohl, J.~Petersen, and K.~H. Maier-Hein,
  ``nnu-net: a self-configuring method for deep learning-based biomedical image
  segmentation,'' {\em Nature methods}, vol.~18, no.~2, pp.~203--211, 2021.

\bibitem{chen2021transunet}
J.~Chen, Y.~Lu, Q.~Yu, X.~Luo, E.~Adeli, Y.~Wang, L.~Lu, A.~L. Yuille, and
  Y.~Zhou, ``Transunet: Transformers make strong encoders for medical image
  segmentation,'' {\em arXiv preprint arXiv:2102.04306}, 2021.

\bibitem{cao2021swin}
H.~Cao, Y.~Wang, J.~Chen, D.~Jiang, X.~Zhang, Q.~Tian, and M.~Wang,
  ``Swin-unet: Unet-like pure transformer for medical image segmentation,''
  {\em arXiv preprint arXiv:2105.05537}, 2021.

\bibitem{ouyang2022self}
C.~Ouyang, C.~Biffi, C.~Chen, T.~Kart, H.~Qiu, and D.~Rueckert,
  ``Self-supervised learning for few-shot medical image segmentation,'' {\em
  IEEE Transactions on Medical Imaging}, vol.~41, no.~7, pp.~1837--1848, 2022.

\bibitem{tang2021recurrent}
H.~Tang, X.~Liu, S.~Sun, X.~Yan, and X.~Xie, ``Recurrent mask refinement for
  few-shot medical image segmentation,'' in {\em Proceedings of the IEEE/CVF
  international conference on computer vision}, pp.~3918--3928, 2021.

\bibitem{kirillov2023segment}
A.~Kirillov, E.~Mintun, N.~Ravi, H.~Mao, C.~Rolland, L.~Gustafson, T.~Xiao,
  S.~Whitehead, A.~C. Berg, W.-Y. Lo, {\em et~al.}, ``Segment anything,'' {\em
  arXiv preprint arXiv:2304.02643}, 2023.

\bibitem{zhang_towards_2023}
Y.~Zhang and R.~Jiao, ``Towards segment anything model ({SAM}) for medical
  image segmentation: A survey,'' {\em arXiv preprint arXiv:2305.03678}.

\bibitem{zhao2019data}
A.~Zhao, G.~Balakrishnan, F.~Durand, J.~V. Guttag, and A.~V. Dalca, ``Data
  augmentation using learned transformations for one-shot medical image
  segmentation,'' in {\em Proceedings of the ieee/cvf conference on computer
  vision and pattern recognition}, pp.~8543--8553, 2019.

\bibitem{ding2021modeling}
Y.~Ding, X.~Yu, and Y.~Yang, ``Modeling the probabilistic distribution of
  unlabeled data for one-shot medical image segmentation,'' in {\em Proceedings
  of the AAAI conference on artificial intelligence}, vol.~35, pp.~1246--1254,
  2021.

\bibitem{wang2020lt}
S.~Wang, S.~Cao, D.~Wei, R.~Wang, K.~Ma, L.~Wang, D.~Meng, and Y.~Zheng,
  ``Lt-net: Label transfer by learning reversible voxel-wise correspondence for
  one-shot medical image segmentation,'' in {\em Proceedings of the IEEE/CVF
  Conference on Computer Vision and Pattern Recognition}, pp.~9162--9171, 2020.

\bibitem{he2022learning}
Y.~He, R.~Ge, X.~Qi, Y.~Chen, J.~Wu, J.-L. Coatrieux, G.~Yang, and S.~Li,
  ``Learning better registration to learn better few-shot medical image
  segmentation: Authenticity, diversity, and robustness,'' {\em IEEE
  Transactions on Neural Networks and Learning Systems}, 2022.

\bibitem{collins1995automatic}
D.~L. Collins, C.~J. Holmes, T.~M. Peters, and A.~C. Evans, ``Automatic 3-d
  model-based neuroanatomical segmentation,'' {\em Human brain mapping},
  vol.~3, no.~3, pp.~190--208, 1995.

\bibitem{lorenzo2002atlas}
M.~Lorenzo-Vald{\'e}s, G.~I. Sanchez-Ortiz, R.~Mohiaddin, and D.~Rueckert,
  ``Atlas-based segmentation and tracking of 3d cardiac mr images using
  non-rigid registration,'' in {\em International Conference on Medical image
  computing and computer-assisted intervention}, pp.~642--650, Springer, 2002.

\bibitem{lotjonen2010fast}
J.~M. L{\"o}tj{\"o}nen, R.~Wolz, J.~R. Koikkalainen, L.~Thurfjell, G.~Waldemar,
  H.~Soininen, D.~Rueckert, A.~D.~N. Initiative, {\em et~al.}, ``Fast and
  robust multi-atlas segmentation of brain magnetic resonance images,'' {\em
  Neuroimage}, vol.~49, no.~3, pp.~2352--2365, 2010.

\bibitem{coupe2011patch}
P.~Coup{\'e}, J.~V. Manj{\'o}n, V.~Fonov, J.~Pruessner, M.~Robles, and D.~L.
  Collins, ``Patch-based segmentation using expert priors: Application to
  hippocampus and ventricle segmentation,'' {\em NeuroImage}, vol.~54, no.~2,
  pp.~940--954, 2011.

\bibitem{holden2007review}
M.~Holden, ``A review of geometric transformations for nonrigid body
  registration,'' {\em IEEE transactions on medical imaging}, vol.~27, no.~1,
  pp.~111--128, 2007.

\bibitem{kennedy2012candishare}
D.~N. Kennedy, C.~Haselgrove, S.~M. Hodge, P.~S. Rane, N.~Makris, and J.~A.
  Frazier, ``Candishare: a resource for pediatric neuroimaging data,'' {\em
  Neuroinformatics}, vol.~10, pp.~319--322, 2012.

\bibitem{di2014autism}
A.~Di~Martino, C.-G. Yan, Q.~Li, E.~Denio, F.~X. Castellanos, K.~Alaerts, J.~S.
  Anderson, M.~Assaf, S.~Y. Bookheimer, M.~Dapretto, {\em et~al.}, ``The autism
  brain imaging data exchange: towards a large-scale evaluation of the
  intrinsic brain architecture in autism,'' {\em Molecular psychiatry},
  vol.~19, no.~6, pp.~659--667, 2014.

\bibitem{wang2020generalizing}
Y.~Wang, Q.~Yao, J.~T. Kwok, and L.~M. Ni, ``Generalizing from a few examples:
  A survey on few-shot learning,'' {\em ACM computing surveys (csur)}, vol.~53,
  no.~3, pp.~1--34, 2020.

\bibitem{dong2018few}
N.~Dong and E.~P. Xing, ``Few-shot semantic segmentation with prototype
  learning.,'' in {\em BMVC}, vol.~3, 2018.

\bibitem{wang2019panet}
K.~Wang, J.~H. Liew, Y.~Zou, D.~Zhou, and J.~Feng, ``Panet: Few-shot image
  semantic segmentation with prototype alignment,'' in {\em proceedings of the
  IEEE/CVF international conference on computer vision}, pp.~9197--9206, 2019.

\bibitem{zhang2020sg}
X.~Zhang, Y.~Wei, Y.~Yang, and T.~S. Huang, ``Sg-one: Similarity guidance
  network for one-shot semantic segmentation,'' {\em IEEE transactions on
  cybernetics}, vol.~50, no.~9, pp.~3855--3865, 2020.

\bibitem{li2021adaptive}
G.~Li, V.~Jampani, L.~Sevilla-Lara, D.~Sun, J.~Kim, and J.~Kim, ``Adaptive
  prototype learning and allocation for few-shot segmentation,'' in {\em
  Proceedings of the IEEE/CVF conference on computer vision and pattern
  recognition}, pp.~8334--8343, 2021.

\bibitem{kwon2021dual}
H.~Kwon, S.~Jeong, S.~Kim, and K.~Sohn, ``Dual prototypical contrastive
  learning for few-shot semantic segmentation,'' {\em arXiv preprint
  arXiv:2111.04982}, 2021.

\bibitem{chen2021semantically}
T.~Chen, G.-S. Xie, Y.~Yao, Q.~Wang, F.~Shen, Z.~Tang, and J.~Zhang,
  ``Semantically meaningful class prototype learning for one-shot image
  segmentation,'' {\em IEEE Transactions on Multimedia}, vol.~24, pp.~968--980,
  2021.

\bibitem{yang2021mining}
L.~Yang, W.~Zhuo, L.~Qi, Y.~Shi, and Y.~Gao, ``Mining latent classes for
  few-shot segmentation,'' in {\em Proceedings of the IEEE/CVF international
  conference on computer vision}, pp.~8721--8730, 2021.

\bibitem{chenrole}
C.-Y. Chen, H.-T. Lin, M.~Sugiyama, and G.~Niu, ``On the role of pre-training
  for meta few-shot learning,'' in {\em Fifth Workshop on Meta-Learning at the
  Conference on Neural Information Processing Systems}.

\bibitem{he2020deep}
Y.~He, T.~Li, G.~Yang, Y.~Kong, Y.~Chen, H.~Shu, J.-L. Coatrieux, J.-L.
  Dillenseger, and S.~Li, ``Deep complementary joint model for complex scene
  registration and few-shot segmentation on medical images,'' in {\em 16th
  European conference on computer vision}, pp.~770--786, Springer, 2020.

\bibitem{xu2019deepatlas}
Z.~Xu and M.~Niethammer, ``Deepatlas: Joint semi-supervised learning of image
  registration and segmentation,'' in {\em International Conference on Medical
  image computing and computer assisted intervention}, pp.~420--429, Springer,
  2019.

\bibitem{balakrishnan2019voxelmorph}
G.~Balakrishnan, A.~Zhao, M.~R. Sabuncu, J.~Guttag, and A.~V. Dalca,
  ``Voxelmorph: a learning framework for deformable medical image
  registration,'' {\em IEEE transactions on medical imaging}, vol.~38, no.~8,
  pp.~1788--1800, 2019.

\bibitem{goodfellow2014explaining}
I.~J. Goodfellow, J.~Shlens, and C.~Szegedy, ``Explaining and harnessing
  adversarial examples,'' {\em arXiv preprint arXiv:1412.6572}, 2014.

\bibitem{kim2021cyclemorph}
B.~Kim, D.~H. Kim, S.~H. Park, J.~Kim, J.-G. Lee, and J.~C. Ye, ``Cyclemorph:
  cycle consistent unsupervised deformable image registration,'' {\em Medical
  image analysis}, vol.~71, p.~102036, 2021.

\bibitem{rueckert1999nonrigid}
D.~Rueckert, L.~I. Sonoda, C.~Hayes, D.~L. Hill, M.~O. Leach, and D.~J. Hawkes,
  ``Nonrigid registration using free-form deformations: application to breast
  mr images,'' {\em IEEE transactions on medical imaging}, vol.~18, no.~8,
  pp.~712--721, 1999.

\bibitem{zheng2021rectifying}
Z.~Zheng and Y.~Yang, ``Rectifying pseudo label learning via uncertainty
  estimation for domain adaptive semantic segmentation,'' {\em International
  Journal of Computer Vision}, vol.~129, no.~4, pp.~1106--1120, 2021.

\bibitem{milletari2016v}
F.~Milletari, N.~Navab, and S.-A. Ahmadi, ``V-net: Fully convolutional neural
  networks for volumetric medical image segmentation,'' in {\em 2016 fourth
  international conference on 3D vision (3DV)}, pp.~565--571, IEEE, 2016.

\bibitem{monai}
``Medical open network for ai (monai).''
\newblock \url{https:/https://github.com/Project-MONAI/MONAI} Accessed Feburary
  19, 2023.

\bibitem{luo2022semi}
X.~Luo, G.~Wang, W.~Liao, J.~Chen, T.~Song, Y.~Chen, S.~Zhang, D.~N. Metaxas,
  and S.~Zhang, ``Semi-supervised medical image segmentation via uncertainty
  rectified pyramid consistency,'' {\em Medical Image Analysis}, vol.~80,
  p.~102517, 2022.

\bibitem{wu2022mutual}
Y.~Wu, Z.~Ge, D.~Zhang, M.~Xu, L.~Zhang, Y.~Xia, and J.~Cai, ``Mutual
  consistency learning for semi-supervised medical image segmentation,'' {\em
  Medical Image Analysis}, vol.~81, p.~102530, 2022.

\end{thebibliography}

\end{document}